\documentclass[%
 reprint,
 amsmath,amssymb,
 aps,
]{revtex4-2}

\usepackage{graphicx}
\usepackage{dcolumn}
\usepackage{bm}
\graphicspath{{Figures/}}

\begin{document}




\title{Collective synchronization of undulatory movement through contact}

\author{Wei Zhou}
\author{Zhuonan Hao}
\author{Nick Gravish}%
 \email{ngravish@ucsd.edu}
\affiliation{%
 Mechanical and Aerospace Engineering,  University of California, San Diego, 92093
}%

\date{\today}



\begin{abstract} 
    Many biological systems synchronize their movement through physical interactions. By far the most well studied examples concern physical interactions through a fluid: beating cilia, swimming sperm and worms, and flapping wings, all display synchronization behavior through fluid mechanical interactions. However, as the density of a collective increases individuals may also interact with each other through physical contact. In the field of ``active matter'' systems, it is well known that inelastic contact between individuals can produce long-range correlations in position, orientation, and velocity. In this work we demonstrate that contact interactions between undulating robots yield novel phase dynamics such as synchronized motions. We consider undulatory systems in which rhythmic motion emerges from time-independent oscillators that sense and respond to undulatory bending angle and speed. In pair experiments we demonstrate that robot joints will synchronize to in-phase and anti-phase oscillations through collisions and a phase-oscillator model describes the stability of these modes. To understand how contact interactions influence the phase dynamics of larger groups we perform simulations and experiments of simple three-link undulatory robots that interact only through contact. Collectives synchronize their movements through contact as predicted by the theory and when the robots can adjust their position in response to contact we no longer observe anti-phase synchronization. Lastly we demonstrate that synchronization dramatically reduces the interaction forces within confined groups of undulatory robots indicating significant energetic and safety benefits from group synchronization. The theory and experiments in this study illustrate how contact interactions in undulatory active matter can lead to novel collective motion and synchronization.
\end{abstract}
\maketitle



The study of oscillations in biological systems have lead to fundamental understanding of the dynamics of coupled oscillators \cite{Winfree2001-ft, Winfree1967-wx}. 
Biological locomotion typically arises from oscillatory movements and groups of living systems can exhibit coupled movement oscillations when interacting. 
For example recent studies have demonstrated that fluid-forces acting between pairs of flagella \cite{Wan2016-tw, Brumley2014-cg, Geyer2013-fv}, arrays of cilia \cite{Brumley2012-lc, Han2018-rr, Gilpin2020-nc}, and even flapping wings \cite{Newbolt2019-gg, Oza2019-xf, Becker2015-xt} can lead to phase and frequency synchronization of oscillatory body movements. 
However, many animal and robot groups operate in close proximity where movements may result in collisions, leading to collective jamming \cite{Gravish2015-qt, Aguilar2018-bu}, disorder-to-order transitions in traffic flow \cite{Buhl2006-qf, Csahok1994-hs}, and synchronization of oscillatory swimming gaits \cite{Yuan2014-xs}. 
In this work we study the phase dynamics of oscillators that are coupled only through intermittent mechanical contact. 
We provide experimental and theoretical evidence that inelastic mechanical collisions between independent oscillators produce a rich array of phase dynamics in contact-coupled systems. 


\begin{figure}[b]
    \centering
    \includegraphics[width=0.95\columnwidth]{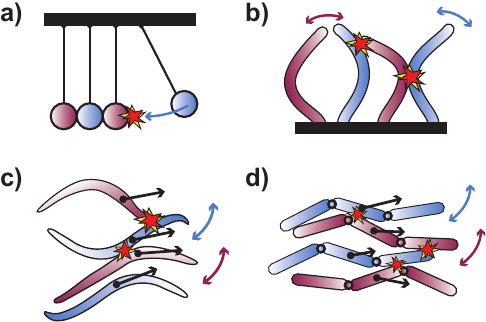}
    \caption{Examples of oscillators that interact through contact. a) The balls of the Newton's cradle toy collide and synchronize to in-phase oscillations. b) Arrays of flapping cilia in close proximity can be driven to synchrony through contact  \cite{Chelakkot2020-fq}. c) The undulatory gait of swimming worms (\textit{C. elegans}) synchronize through contact interactions \cite{Yuan2014-xs}. d) Simple three-link ``Purcell-swimmer'' robots similarly synchronize their gaits through contact as demonstrated in this work.}
    \label{fig:01_examples}
\end{figure}

Synchronization in biological systems can be observed across all scales---from genetic oscillators within cells \cite{Danino2010-qf,Tayar2017-pe} to collective animal groups within habitats \cite{Mirollo1990-dw,Sarfati2020-za,Strogatz2000-rs}.
While synchronization is observed across a wide variety of different systems ultimately it requires two fundamental properties \cite{Pikovsky2003-rf}: (1) perturbations to the phase of each oscillator neither grow or decay, and (2) oscillator interactions can influence phase. 
Many mechanical systems possess both such properties, for example the original pendulum clocks of Huygens \cite{Bennett2002-ps} exhibit (1) autonomous oscillations that (2) interact through structural motion. 
In the context of undulatory locomotion there are two main archetypes for autonomous oscillations \cite{Marder2001-ev}: central pattern generators that provide an adaptive global ``clock'', and reflexive oscillators that generate spontaneous oscillations through local feedback. 
Critically both modalities incorporate environmental and proprioceptive feedback.
Many abstractions of these circuits exist \cite{Ijspeert2008-ty, Matsuoka1985-dv, Kopell1988-xy} and one common model is the phase-oscillator which oscillates at a constant frequency $\omega$ and can be augmented with sensory feedback.  


Collectives that interact through contact have been extensively studied in the soft-matter fields, such as active matter and granular materials. 
Inert systems that interact through contact such as granular materials exhibit novel nonlinear phenomena such as inelastic collapse \cite{Topic2019-oq, McNamara1994-rv}, jamming \cite{Behringer2019-nd}, and transitions between fluid and solid states \cite{Jaeger1996-fw}. 
However, granular materials require external driving forces to stay in motion.
In contrast active matter systems generate spontaneous movement through internal energy reservoirs and external interactions with the group \cite{Ramaswamy2010-cz}. 
Studies of self-propelled ``dry'' active matter have demonstrated that inelastic collisions are responsible for the collective motion and long-range order in dynamical states such as flocking, jamming, and phase separation (see \cite{Shaebani2020-mr} for a review). 
Recent experiments and simulation of undulatory active matter systems such as swimming sperm \cite{Yang2008-hj,Yang2010-ps}, and reciprocating robots \cite{Chvykov2021-jx,Savoie2019-gx}, have demonstrated that contact interactions can lead to novel spatial ordering. 
However, the explicit ability for these mobile systems to adapt phase and synchronize through contact is unknown. 
In this manuscript we study an active matter system of undulatory robots and demonstrate that inelastic mechanical collisions produce a rich dynamics of collective behavior through contact-coupling alone. 

\begin{figure*}[t]
    \centering
    \includegraphics[width=1 \linewidth]{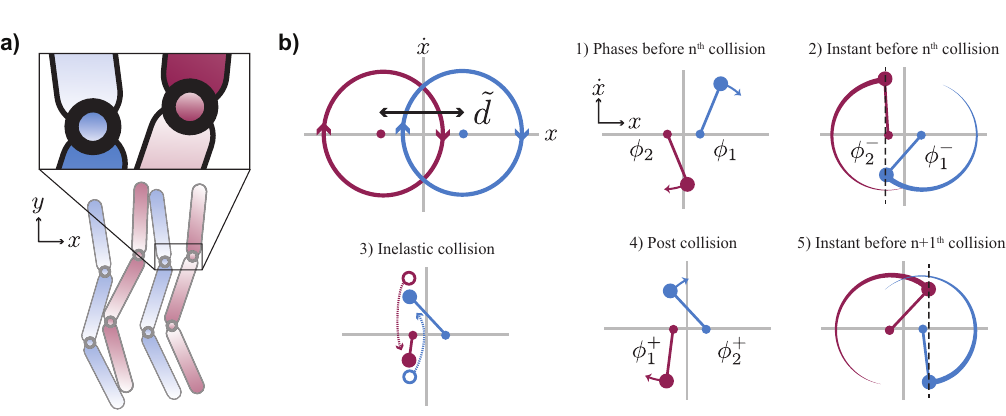}
    \caption{A phase oscillator model for contact mediated synchronization of undulatory gaits. a) Undulatory motion is generated through periodic bending of body elements at joints. b) We envision that the motion of the body in the lateral direction ($x$) is governed by a phase oscillator that produces harmonic motion.  The evolution of the collision model is shown in steps 1-5. 1) Oscillators are initially at phase difference $\Delta = \phi_2 - \phi_1$. 2) Oscillators will collide when $x_1 = x_2$. 3) During a collision the velocities are instantaneously updated according to an inelastic collision law and the phase difference changes. 4) Immediately post-collision the oscillators continue evolving until 5) they collide again resulting in a new post-collision phase difference.}
    \label{fig:02_model1}
\end{figure*}

As a first example of contact-coupled synchronization we introduce the Newton's cradle toy (Fig.~\ref{fig:01_examples}). 
Newton's cradle is a series of metal balls mounted on wires so that they each undergo pendular motion. 
When one ball is allowed to fall under pendular motion and collide with the group, energy is transferred through collisions (with some energy loss) and the ball on the other end will rotate upwards. 
A less appreciated aspect of this process is that as time evolves energy is lost due to collisions and eventually the system settles into a state where all of the pendulums are oscillating in phase and in continuous contact. 
This is a simple example of a contact-coupled dynamical system in which the pendulum are initially out of phase, but through repeated collisions and energy loss the system is driven to a synchronous oscillating state. 

In active oscillating systems energy loss through dissipation or collisions can be compensated for by energy input, thus exhibiting limit-cycle oscillations \cite{Strogatz2000-rs}. 
The oscillatory movements of some biological systems can be considered as limit-cycle oscillators \cite{Buchli2006-nr,Ijspeert2008-ty} prompting our interest in the phase dynamics of active oscillatory systems that interact through contact (Fig.~\ref{fig:01_examples}). 
We consider simplified representations of biological systems that move through undulation: our experimental robots use rotary joints and have rigid links. 
To allow the robots to evolve in undulatory phase we use a simple autonomous nonlinear oscillator to drive sinusoidal motions of the robots. 

In the following sections we study how oscillatory systems can achieve synchronization when they interact through contact. In section I we present a simple theoretical model of contact synchronization and we analyze the steady-state modes and their stability through a contact-to-contact iterated map. In section II we introduce a simple experiment to examine how two robot joints can synchronize through mechanical contact and we compare with the theoretical model. To understand how contact interactions may apply to larger groups we perform simulations of 1D lattices and study their dynamics in section III. In the last two sections we perform experiments and simulations on simple three-link robots that interact through collisions. We demonstrate that the in-phase synchronization predicted by our phase model, and observed in our first experiments, is observed in robot groups (section IV). To motivate why synchronization is beneficial in mobile groups we measure contact forces between robots and compare between the synchronous and asynchronous states. When robot joint oscillation is driven through time-dependent sinusoidal control the contact forces are orders of magnitude larger than when the robots are allowed to synchronize.

\section{A model of synchronization through contact}
\label{sec:theory}
We begin by studying a simple model of two phase oscillators that represent body-bending elements, or the joints of undulatory robots (Fig.~\ref{fig:02_model1}a).
We consider that undulatory motion is generated according to the phase oscillator equation, $\dot{\phi} = 1$. 
The oscillator phase governs the lateral position of the undulating body-element, such that $x_i = A \cos(\phi_i)$ is the lateral distance from the body center-line, and $\dot{x}_i = -A \sin(\phi_i)$ is the lateral speed  (Fig.~\ref{fig:02_model1}b). 
When two body elements are in close proximity they will come into contact when the following condition is met $A \cos(\phi_1) - A \cos(\phi_2) = d$, where $d$ is the separation distance of the central axis of the two agents. 
We introduce the normalized separation distance, $\tilde{d} = \frac{d}{2A}$, such that only when the condition $\tilde{d} \leq 1$ will oscillators be able to contact. 

The contact condition thus becomes 
\begin{align}
    \cos(\phi_2) - \cos(\phi_1) = 2 \tilde{d} 
    \label{eqn:collision_condition}
\end{align}
When the oscillator pair collides they each have a velocity of $\dot{x}_i^{-} = -A \sin(\phi_i^{-})$ where superscripts $\pm$ denote before ($-$) and after ($+$) collision variables (Fig.~\ref{fig:02_model1}b).
We model the collision as an inelastic process with coefficient of restitution $r$ such that $\dot{x}_1^{+} - \dot{x}_2^{+} = -r\left(\dot{x}_1^{-} - \dot{x}_2^{-}\right)$. 
Combining the inelastic collision model with conservation of momentum, $\dot{x}_1^{+} + \dot{x}_2^{+} = \dot{x}_1^{-} + \dot{x}_2^{-}$, yields the following post-collision velocities (we assume equal masses)
\begin{align}
    \label{eqn:coeffrest1}
    \dot{x}_1^{+} = \frac{1}{2} \left[\left(1 - r \right) \dot{x}_1^{-} + \left(1 + r \right) \dot{x}_2^{-}\right] \\
    \label{eqn:coeffrest2}
    \dot{x}_2^{+} = \frac{1}{2} \left[\left(1 - r \right) \dot{x}_2^{-} + \left(1 + r \right) \dot{x}_1^{-}\right]
\end{align}

When the oscillators collide they instantaneously change their phase due to the velocity change (Fig.~\ref{fig:02_model1}b). 
The oscillator phase is represented in the phase-plane as the clockwise angle from the positive $x$ axis to the instantaneous coordinate $(x, \dot{x})$. 
Thus, the phases before and after a collision can be represented by the following equation 
\begin{align}
    \label{eqn:post_collide_phase}
    \phi_i^{\pm} = \mathrm{atan} \left[-\frac{\dot{x}_i^{\pm}}{x_i^{\pm}}\right]
\end{align}
The negative sign accounts for the fact that the rotation direction is in the clockwise direction. 

\begin{figure*}[t]
    \centering
    \includegraphics[width=1 \linewidth]{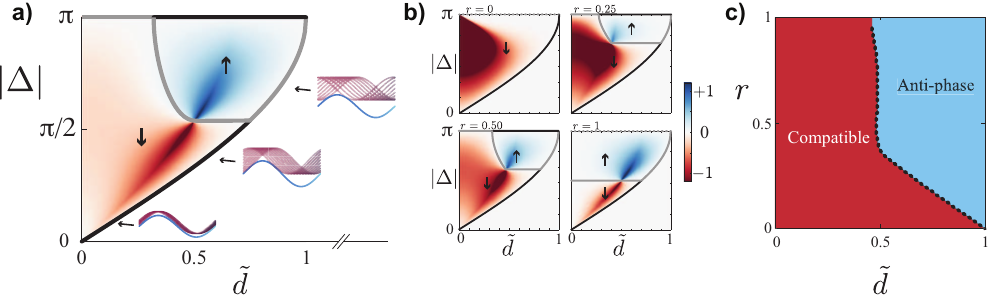}
    \caption{
        The evolution of the phase difference is captured by the phase oscillator model. a) For an initial phase difference, $\Delta$, and separation distance, $\tilde{d}$, a collision will induce a change in $\Delta$. The heatmap shows the collision to collision phase difference, $|\Delta^{(n+1)}| - |\Delta^{(n)}|$ at each $(\tilde{d}, \Delta)$ location (Eqn.~\ref{eqn:main_compatibility}). States in the red region result in a decrease in the absolute phase difference, while states in the blue region increase in phase difference. Black lines are stable fixed points, solid gray lines are unstable, and dashed gray lines are marginally stable. The lower black curve is the compatibility fixed point (Eqn.~\ref{eqn:compatibility}). The coefficient of restitution is $r = 0.67$. The inset shows three different separation distances and the range of ``compatible'' phase differences that can exist without collision. b) Collision to collision phase change behavior at four different restitution coefficients. The black lines are stable fixed points, the gray lines are unstable fixed points, and dashed gray lines are neutrally stable. c) The average steady-state from random initial phases as a function of separation distance and restitution coefficient.}
    \label{fig:03_model2}
\end{figure*}

We seek to understand the asymptotic behavior of the phase difference, $\Delta = \phi_2 - \phi_1$. 
In systems with continuous coupling this often amounts to demonstrating that $\dot{\Delta} = 0$ \cite{Pikovsky2003-rf}.
However, since this system consists of repetitive collision events the phase difference, $\Delta$, is constant in between collisions and changes instantaneously during a collision. 
Thus, we will derive the iterated map that takes the pre-collision phase difference of the $(n)^{\text{th}}$ collision to the pre-collision phase difference of the $(n+1)^{\text{th}}$ collision, $\Delta^{(n+1)} = f \left(\Delta^{(n)}\right)$. 
We represent the phase difference of the $n^{\text{th}}$ collision as $\Delta^{(n)}$ where we have dropped the superscript $+$ for notation convenience. 
To derive $f \left(\Delta^{(n)}\right)$ we take the following steps: 1) solve for $\phi_1$ and $\phi_2$ at collision which gives $\Delta^{(n,-)}$, 2) apply the velocity update rule for the inelastic collision, 3) determine the post-collision phases for the oscillators. 
Since $\omega$ is the same between each oscillator the $n^{\text{th}}$ post-collision phase difference is exactly the same as the $n+1^{\text{th}}$ pre-collision phase difference and thus $\Delta^{(n,+)} = \Delta^{(n+1,-)}$. 
We have now determined the function that generates $\Delta^{(n+1,-)}$ from $\Delta^{(n,-)}$ and we can drop the $\pm$ superscripts yielding $\Delta^{(n+1)} = f \left(\Delta^{(n)}\right)$. This results in the collision-to-collision return map 
\begin{widetext}
\begin{align}
    \label{eqn:main_compatibility}
    \Delta^{(n+1)} = \mathrm{atan} \left[\frac{\sin (\kappa) \cos (\frac{\Delta^{(n)}}{2})- r \cos (\kappa) \sin (\frac{\Delta^{(n)}}{2})}{\cos (\kappa) \cos (\frac{\Delta^{(n)}}{2})-\sin (\kappa) \sin (\frac{\Delta^{(n)}}{2})}\right] - \mathrm{atan} \left[\frac{\sin (\kappa) \cos (\frac{\Delta^{(n)}}{2}) + r \cos (\kappa) \sin (\frac{\Delta^{(n)}}{2})}{\cos (\kappa) \cos (\frac{\Delta^{(n)}}{2}) +\sin (\kappa) \sin (\frac{\Delta^{(n)}}{2})}\right]
\end{align}
\end{widetext}
where we have defined $\kappa = \mathrm{asin}\left(\tilde{d}\csc\left(\frac{\Delta}{2}\right)\right)$.

The collision-to-collision return map allows us to examine the asymptotic behavior and dynamics of synchronization for contact coupled oscillators. 
We first examine the fixed points of the map, $\Delta^{*} = f \left(\Delta^{*}\right)$.
The return map for $0 < r \leq 1$ exhibits three fixed points as a function of separation. Two of the fixed points exist independent of the coefficient of restitution
\begin{align}
    \label{eqn:compatibility}
    \Delta^* & = 2\: \mathrm{asin}(\tilde{d})  \\ 
    \label{eqn:anti-phase}
    \Delta^* &= \pi
\end{align}
while the third fixed point depends on $r$ and must be solved numerically.
We define the first fixed point (Eqn.~\ref{eqn:compatibility}) as the compatibility curve, because it defines the maximum phase difference between two oscillators separated by $\tilde{d}$ before they will collide (see inset Fig.~\ref{fig:03_model2}a).
When oscillators are at the compatibility fixed point they will repeatedly make grazing contact with each other.
The compatibility curve actually determines the boundary of an entire set of fixed points for these oscillators, since if the phase difference, $|\Delta^*| < 2\: \mathrm{asin}(\tilde{d})$, the oscillators will never contact each other and thus $\Delta$ will never change. 
The second fixed point (Eqn.~\ref{eqn:anti-phase}) is an anti-phase oscillation. 
We show the fixed points in Figure~\ref{fig:03_model2}a for $r = 0.67$ where the lower branch is the compatibility fixed point, and the upper branch is the anti-phase fixed point.

When the separation distance is zero ($\tilde{d} = 0$) the compatibility fixed point corresponds to perfect in-phase synchronization, $\Delta^* = 0$ and the return map dramatically simplifies to $\Delta^{(n+1)} = -2\: \mathrm{atan} \left(r \tan \left( \frac{1}{2} \Delta^{(n)}\right)\right)$.
This equation can be solved recursively to generate the phase difference of the $n^{\mathrm{th}}$ collision as a function of any initial condition ($\Delta^{(0)}$)
\begin{align}
    \Delta^{(n)} = -2 \: \mathrm{atan} \left((-r)^{n} \tan \left( \frac{1}{2} \Delta^{(0)}\right)\right)
    \label{eqn:d_0}
\end{align}
and we clearly see that for large $n$ the phase difference converges to $\Delta^{*} = 0$.

Equation \ref{eqn:d_0} highlights the importance of inelastic collisions in the synchronization process for contact coupled oscillators. 
The coefficient of restitution, $r$, governs the rate of convergence to the synchronization fixed point for $\tilde{d}$. 
The linear stability of fixed points in the collision-to-collision map are determined by the condition $|f'(0)| < 1$ where prime denotes the derivative with respect to $\Delta^{(n)}$.
For the $\Delta^* = 0$ fixed point the stability is $f'(0) = -r$ again highlighting the importance of inelasticity in the synchronization process. 
Thus, because inelastic interactions always generate energy loss ($0 < r < 1$) the system is guaranteed to reach phase synchronization when $\tilde{d} = 0$. 

\begin{figure*}[t]
    \centering
    \includegraphics[width=1\linewidth]{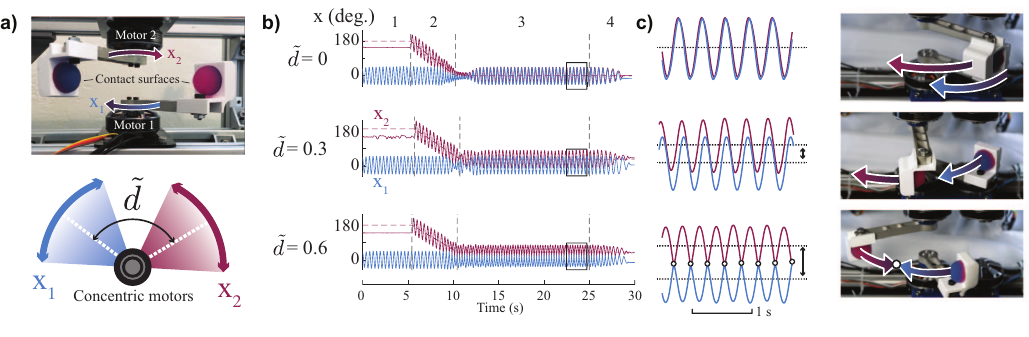}
    \caption{Experimental validation of synchronization between undulatory robot joints. a) Two motors mounted on a concentric axis are actuated as phase oscillators. The oscillators interact through inelastic collisions when their rotation angle is equal. b) Data from three separation distances. Robot joints are initially oscillated out of contact to achieve steady-state behavior (Phases 1 and 2) and are slowly brought into contact (Phase 2) to their final fixed distance, $\tilde{d}$ (Phase 3), until the experiment is over (Phase 4). c) At low $\tilde{d}$ joints synchronize, at intermediate $\tilde{d}$ joints oscillate with compatible phases and don't contact, and at large $\tilde{d}$ joints collide in anti-phase synchronization. Images are from SI Video 1.}
    \label{fig:04_experiment1}
\end{figure*}
 
To analyze the time evolution of the system when $0 < \tilde{d} < 1$ we will construct the basins of attraction for the fixed points by calculating the phase change behavior after a single collision event,  $g(|\Delta^{(n)}|) = |\Delta^{(n+1)}| - |\Delta^{(n)}|$. 
In Figure~\ref{fig:03_model2}a-b we plot $g(|\Delta^{(n)}|)$ and denote with arrows and color the flow direction of the compatible (down arrow, red) and anti-phase (up arrow, blue) basins. 
We observe that for each $r$ there is a critical $\tilde{d}_c$ below which all initial phase differences are attracted to the compatible state.
However, for larger $\tilde{d}$ the anti-phase basin causes states that start with large $|\Delta|$ to evolve to anti-phase synchronization (blue regions in Fig.~\ref{fig:03_model2}a,b).

To analyze the behavior of the anti-phase fixed point (Eqn.~\ref{eqn:anti-phase}) we similarly construct the basin of attraction and linear stability. 
Since the return map at $\Delta^* = \pi$ has a continuous first derivative we can compute the linear stability of this point.
Evaluating the derivative we find 
\begin{align}
    \label{eqn:antiphase_stability}
    f'(\pi) = -\frac{r\left(d^{2}-1\right)+d^{2}}{r^{2}\left(d^{2}-1\right)-d^{2}}
\end{align} 
which yields the critical separation distance, $\tilde{d}_c = \sqrt{\frac{r ( r - 1 ) }{r^2 - r - 2}}$.
When $\tilde{d} > \tilde{d}_c$ anti-phase oscillations switch from unstable to stable. 
However as $r \rightarrow 0$ the range of $\tilde{d}$ where $|f'(\pi)| < 1$ becomes vanishingly small as $f'(\pi)$ converges to $f'(\pi) = 1$ for infinitesimal $\tilde{d}$.
The overall influence of $r$ and $\tilde{d}$ can be understood by averaging the collision-to-collision phase change across all initial phases, highlighting that for modest $r \approx 0.5$ and above the average steady-state behavior is evenly divided between the anti-phase and compatible states (Fig.~\ref{fig:03_model2}c).

In this section we have proven that a simple model of phase oscillators interacting through intermittent inelastic collisions can produce a rich range of dynamical behavior. 
We observe in-phase synchronization for small separation distances, and anti-phase synchronization for larger distances.
Furthermore, this system admits a continuum of ``fixed points'' when the phase difference is below the compatibility line, in which case the oscillators are completely uncoupled and do not contact.
In the next sections we will demonstrate in experiment and simulation that the pair-wise interactions of contact-coupled oscillators leads to rich collective behaviors.

\section{Synchronization of robot joints in experiment}
\label{sec:experiment1}
 
\begin{figure*}[t]
    \centering
    \includegraphics[width=.95\linewidth]{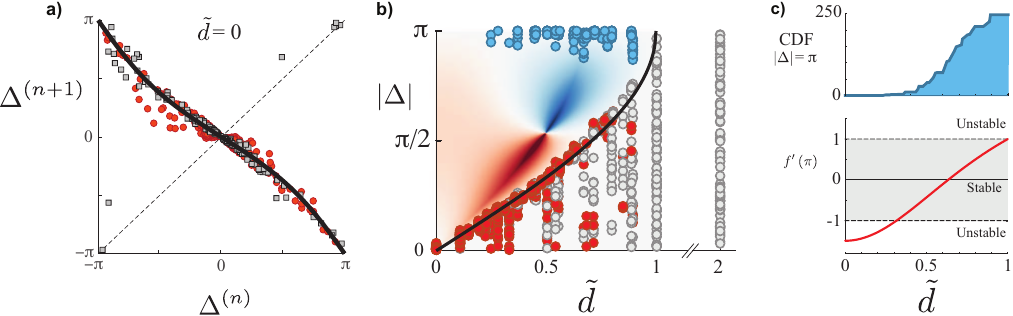}
    \caption{Experiment results. a) Collision return map from two separate experiment methods (over 100 experiments for each). Squares are from experiments in which limit cycles are slowly moved together to $\tilde{d} = 0$, circles are experiments in which oscillators are initialized with $\tilde{d} = 0$. b) Steady-state phase difference ($\Delta$) versus separation distance ($\tilde{d})$ from 1312 experiments. The black line represents the compatibility curve (Eqn.~\ref{eqn:compatibility}). Gray and red circles are points that reached a steady-state configuration in which they no longer collide. Red circles started with $|\Delta|$ above the compatibility curve and evolved downwards to the compatible state, while gray circles represent initial conditions below the compatibility line. Blue points are states that evolved to stable anti-phase oscillations in which the oscillators repeatedly collide head-on. Heatmap is the model prediction from the collision-to-collision return map for $r = 0.67$ (Eqn.~\ref{eqn:main_compatibility}). Far right points at $\tilde{d}=2$ are control experiments. c) Experimental observation of anti-phase oscillations coincide with onset of anti-phase stability in model. Top is the cumulative distribution of observations of anti-phase oscillations versus $\tilde{d}$. Bottom is the stability eigenvalue of anti-phase behavior for $r = 0.67$ (Eqn.~\ref{eqn:antiphase_stability}).}
    \label{fig:05_experiment2}
\end{figure*}

To validate the model introduced in the previous section we performed experiments with two oscillating motors that interact through collisions (Fig.~\ref{fig:04_experiment1}a). Each brushless DC motor (Quanum 5250) represents the joint of a robot and is actuated under closed-loop torque control.
Rigid 9~cm long aluminum links and viscoelastic bumpers were attached to both motors (Fig.~\ref{fig:04_experiment1}a).
We measured the experimental coefficient of restitution of the system to be $r = 0.67$ (SI~Fig.~\ref{fig:SI_restitution}). A capacitive encoder attached to the motor shafts provided angular position measurements at a resolution of 8192 counts per revolution, which is $0.044^\circ$ (AMT102, CUI Devices). An ODrive brushless DC motor controller (ODrive robotics) provides closed loop torque control for both motors individually. 

We consider the joint rotation angle and rotational velocity as the position and velocity variables of our phase oscillator, ($x, \dot{x}$).
In order to actuate these motors as phase oscillators we controlled the motor torque (at a rate of 300~Hz) using the following equation 
\begin{align}
    \tau_i = -k x_i + \left(c - \mu x_i^2\right) \dot{x}_i
    \label{eqn:vdp}
\end{align}
where $x_i$ is the relative angular displacement from the neutral angle and $i$ refers to the oscillator. 
We assume the motor internal damping and friction are small and the systems inertia ($I$) is the same for both motors, such that $I \ddot{x}_i = \tau_i$. 
Note there is no coupling between the motors in equation~\ref{eqn:vdp}, the only interactions are through inelastic collisions. 

The motor actuation Equation~\ref{eqn:vdp} represents a generic form of the Van der Pol oscillator which generates sinusoidal oscillations with constant phase speed ($\dot{\phi}$) for weak nonlinearity \cite{Strogatz2000-rs}.
Thus, this choice of actuation enables the robot joints to oscillate sinusoidally with constant phase velocity consistent with our phase oscillator model in the previous section. 
The position and velocity feedback terms in Equation~\ref{eqn:vdp} enable the actuator to instantaneously respond to collision-induced velocity changes also consistent with our model assumptions.
The actuation parameters of Equation~\ref{eqn:vdp} were chosen such that the oscillators had natural frequencies of $\omega_1 = 2.61 \pm 0.04$~Hz, and $\omega_2 = 2.63 \pm 0.03$~Hz and amplitudes of $A_1 = 44.4 \pm 0.9$~degrees, and $A_2 = 44.3 \pm 1.6$~degrees. For the purposes of analysis and variable definitions we assume equal amplitudes between the oscillators.

To study the phase dynamics between the two colliding oscillators we set up steady limit-cycle oscillations with the systems initially separated by a large neutral position, $\tilde{d} = 2$. 
The lower link was allowed to oscillate and after a random time in the range of 5-7 seconds the upper link was perturbed to limit cycle oscillations. 
This random wait time set a random initial phase difference between the two oscillators. 
Once both links were oscillating at steady-state we slowly moved the neutral position of the second oscillator to the prescribed separation $\tilde{d}$ for that experiment. 
Once the oscillators were at the appropriate $\tilde{d}$ we continued the experiment for 15 seconds until reducing the amplitude and stopping.
We measured the oscillator positions and velocities throughout the experiment (Fig.~\ref{fig:04_experiment1}b) and computed collisional information including the phase difference before each collision, $\Delta^{(n)}$. 
In total we performed 1312 experiments over a range of separation distances where collisions were possible, $\tilde{d} \in [0, 1]$, and a control separation distance $\tilde{d} = 2$ to rule out any coupling through the structure. 
In Figure~\ref{fig:04_experiment1}b-c and supplementary video 1 we show experiments from three $\tilde{d}$ showing in-phase synchronization ($\tilde{d} = 0$), compatibility ($\tilde{d} = 0.3$), and anti-phase synchronization ($\tilde{d} = 0.6$).

\begin{figure*}[t]
    \centering
    \includegraphics[width=1\linewidth]{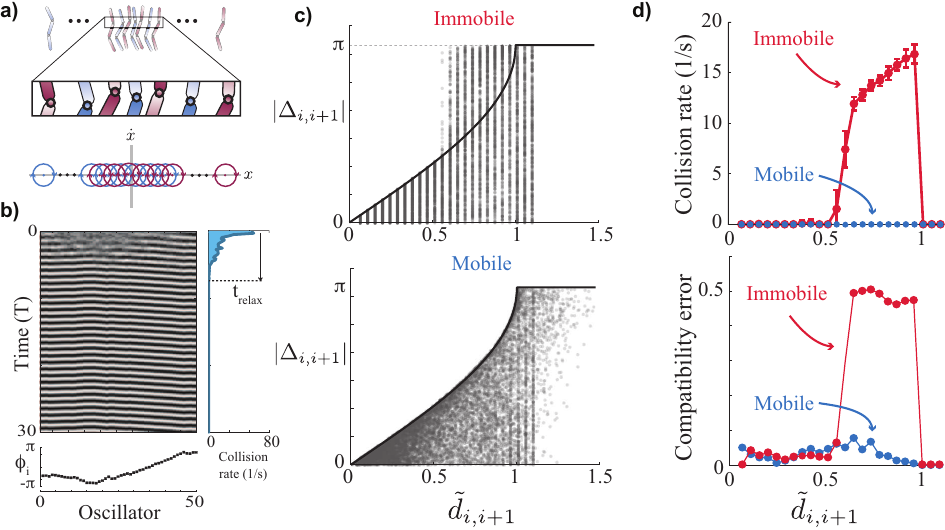}
    \caption{Phase dynamics for oscillator lattices. a) Phase dynamics of undulatory robots are modeled as a one-dimensional lattice with interactions occurring along the lateral direction. Body undulations between neighboring robots can lead to contact. b) The spatio-temporal evolution of phase for an oscillator group (50 oscillators, $\tilde{d} = 0.14$, $\beta = 0$). Bottom plot shows final phases. Right plot shows the rate of collisions and the relaxation time, $t_{relax}$ after which no collisions occur. c) Final phase difference and separation distance between adjacent oscillators. Top shows phase behavior for oscillators with an immobile base ($\beta = 0$). Bottom shows results for mobile oscillators ($\beta = 0.1$) in which the equilibrium position moves as a result of collisions. d) The collision rate and compatibility error versus $\tilde{d}.$ Mobile oscillators always evolve to states with no collisions (top) and good agreement with the compatibility equation (bottom).}
    \label{fig:06_manyoscillators}
\end{figure*}

We first compare the model predictions and experiment for the $\tilde{d} = 0$ return map (Eqn.~\ref{eqn:d_0}).  
In experiment the oscillators synchronized phases through repeated collision events eventually reaching a final synchronized state where the oscillators moved together in or near contact (Fig.~\ref{fig:04_experiment1}b-c). 
The experimental collision-to-collision return map showed consistent in-phase synchronization as predicted by Eqn.~\ref{eqn:d_0} from all initial conditions (Fig.~\ref{fig:05_experiment2}a).
To rule out the effect of slowly bringing the oscillators together (Phase 2 in Fig.~\ref{fig:04_experiment1}b) we performed a second set of experiments in which both oscillators began at $\tilde{d} = 0$ and random initial phase. 
We observed good agreement in the collision-to-collision return map between both experimental methods.
The model and experimental return map exhibited excellent agreement indicating that the phase-oscillator model is able to capture relevant phase dynamics of this system (Eqn.~\ref{eqn:d_0}; black curve in Fig.~\ref{fig:05_experiment2}a). 
It is important to note here that their are no fitting parameters in the model. 
Since the phase dynamics are evaluated from collision-to-collision we do not need to match frequencies or amplitude between experiment and model.
The prediction only requires knowledge of one parameter, the coefficient of restitution $r$ which can easily be measured.

We next compare the steady-state $\Delta$  across the full experimental range of $\tilde{d}$  (Fig~\ref{fig:05_experiment2}b). 
Comparison of the theoretical compatibility curve (Fig.~\ref{fig:05_experiment2}b, solid line) and the experimental data indicates good agreement between the phase oscillator model and observation. We observe that initial phases that start in the compatible state will continue to stay there (gray circles, Fig.~\ref{fig:05_experiment2}b), and initial phases that start outside of the compatible state may either evolve to anti-phase oscillations or compatibility depending on initial conditions.
The red circles in Figure~\ref{fig:05_experiment2}b show initial conditions that began above the compatibility line and evolved to the compatible state.
Blue circles represent initial conditions that began above the compatibility line and evolved to the anti-phase state (Fig.~\ref{fig:05_experiment2}b).

The anti-phase state consisted of the two oscillators repeatedly colliding with each other (see supplementary video 1) in a rather violent manner which lead to broken components on more than one occasion.
The anti-phase state observed in experiment was found to be remarkably stable and able to resist manual perturbations consistent with the stability calculations in Section~\ref{sec:theory}. 
In one experiment we observed the two oscillators remain in the anti-phase state for over 12 hours until we eventually halted the experiment. 
The return map allows us to predict when anti-phase oscillations become stable (Eqn.~\ref{eqn:antiphase_stability}). 
In Fig.~\ref{fig:05_experiment2}c we compare the cumulative observations of anti-phase oscillations and the linear stability calculation (Eqn.~\ref{eqn:antiphase_stability}).
Once again we find exceptional agreement between the model and experiment: as the anti-phase fixed point in our model becomes stable we begin observing anti-phase oscillations in experiment.

\section{Collective behavior of mobile and stationary oscillator groups}
\label{sec:1Dsims}

We next seek to understand whether contact interactions among larger groups can yield similar synchronization and phase dynamics as the robot-pair experiments. 
We consider the lateral dynamics of arrays of mobile cilia and groups of swimming worms as a one-dimensional lattice where nearest-neighbor collisional interactions occur along the direction of body undulations (Fig.~\ref{fig:06_manyoscillators}a).
To simulate the dynamics of mobile and fixed systems we allow the neutral position of each oscillator to move in response to a collision.
Immediately after a collision we update the neutral positions of the colliding oscillators according to the equation $\delta_i = \beta \left(\dot{x}_j - \dot{x}_i \right)$ where $\delta_i $ is the neutral position change of the $i^{\text{th}}$ oscillator and $\beta$ is the magnitude of collision-induced change. 
When $\beta = 0$ the system base is immobile (as in arrays of cilia) while non-zero $\beta$ allows for oscillators to repel each other through collisions.
To confine the oscillator group to a fixed linear distance we set $\beta = 0$ for the left ($i = 0$) and right ($i = N$) oscillators in the $N$-oscillator lattice. 
We perform numerical simulations of the one-dimensional oscillator lattice over varied initial neutral positions spanning $\tilde{d}_{i,i+1} \in [0.06, 1.2]$. 
We simulated 50 oscillators initialized at random phases and observe the phase dynamics, collision rate, and neutral position of the group over time.

When the oscillator lattice is initiated in close proximity (small $\tilde{d}_{i,i+1}$) the oscillators rapidly converge to a compatible state through collisions (Fig.~\ref{fig:06_manyoscillators}b) in both the immobile and mobile cases.
Collisions between oscillators initially occurred due to the random incompatible phases and over time the collision rate decreased ultimately halting after a time, $t_{relax}$, for small $\tilde{d}_{i,i+1} < 0.5$.
Once all oscillators are in the compatible state they will stay there indefinitely unless perturbed. 

To quantitatively compare the oscillator lattice results with the theoretical model and experiments from the previous section we measured the nearest-neighbor phase difference, $|\Delta_{i, i+1}|$, and nearest-neighbor separation distance of the neutral position, $\tilde{d}_{i,i+1}$. 
Examining the relationship between phase difference and spatial separation reveals a fundamental difference between mobile and immobile systems (Fig.~\ref{fig:06_manyoscillators}c).
Immobile oscillator lattices show good agreement between the theoretical predictions and simulation for $\tilde{d}_{i,i+1} < 0.5$.
However when separation distance was large ($\tilde{d}_{i,i+1} > 0.5$) the collisions never stopped and the oscillator groups never entirely reached the compatible state (Fig.~\ref{fig:06_manyoscillators}d; top). 
We measured the collision rate over the last $\approx$~70 periods of oscillation and observe a sharp rise in non-zero steady-state collisions when $\tilde{d}_{i,i+1} > 0.5$ for the immobile system.
This is supplemented by the large cluster of points above the compatibility curve in Figure~\ref{fig:06_manyoscillators}b for the immobile case.
To characterize this deviation from model prediction we calculated the fraction of nearest-neighbor pairs that were above the compatibility curve and plot this compatibility error in Figure~\ref{fig:06_manyoscillators}d.
The immobile base simulations exhibited a large compatibility error and persistent colliding among the group when $\tilde{d} > 0.5$.

The immobile system's deviation from the compatibility curve is easily understood from the pair-wise dynamics of oscillators modeled and studied in the previous sections.
At larger separation distances the collision-to-collision phase change causes oscillator pairs to increase in phase difference.
This phase repulsion is what leads to the stable anti-phase mode in the pair experiment. 
However, in larger groups the interior oscillators have a left and right neighbor and thus experience phase repulsion from both of these neighbors which is not able to relax in simulation. 
These results are in agreement with observations from lattices of locally-coupled Kuramoto oscillators in which repulsive phase interactions have been demonstrated to generate asynchronous collective states \cite{Tsimring2005-la}.

\begin{figure}[t]
    \centering
    \includegraphics[width=1\linewidth]{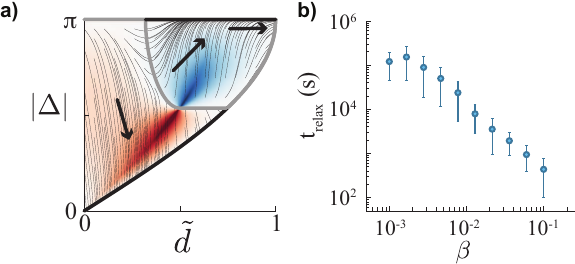}
    \caption{Collision induced mobility allows groups to reach compatibility. a) Phase and spatial evolution of mobile oscillator groups. Collisions result in an increase in separation distance and thus the system evolution tends towards larger $\tilde{d}$ and there is no longer a stable anti-phase state. b) The system relaxation time versus mobility coefficient, $\beta$, for $\tilde{d} = 0.78$. As $\beta$ decreases relaxation time increases. $\beta = 0$ coincides with the immobile simulation in which case the system evolves to stable anti-phase behavior.}
    \label{fig:07_manyoscillators_stats}
\end{figure}

In contrast to the immobile system, oscillators that were able to move in response to collisions always relaxed to the compatible state. 
The phase and spatial values clustered at or below the compatibility curve (Fig.~\ref{fig:06_manyoscillators}c) and exhibited low numbers of collisions and low compatibility error in steady-state (Fig.~\ref{fig:06_manyoscillators}d).
The small but non-zero compatibility error for the mobile system is likely due to the assumptions of pure sinusoidal motion in the theory, compared to the slight deviation in sinusoidal behavior that Equation~\ref{eqn:vdp} generates.
The deviation from compatibility in the mobile system was still small and clustered on or just above the compatibility line.

\begin{figure*}[t]
    \centering
    \includegraphics[width=.95\linewidth]{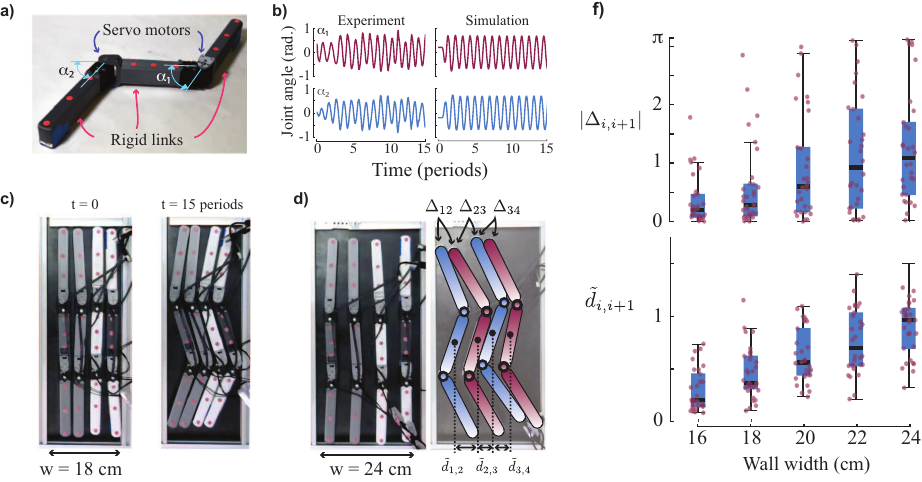}
    \caption{Three-link robot experiments demonstrating synchronization through collisions. a) A simple robot with three rigid links and two servo motors generates undulatory motion. b) The motion of the joints when controlled as a phase oscillator are shown for the experimental system (left) and a simulation (right). c) Experiments with four robots in a controlled width channel demonstrate synchronization of movement over time. d) We measured the phase difference, $\Delta_{i, i+1}$, and lateral separation, $\tilde{d}_{i,i+1}$, between neighboring robots. f) Phase difference and separation distance between robots after 15 periods of oscillation from five different wall width experiments (10 trials each). Boxplots show 25-75\% confidence intervals (blue box) and median (horizontal black line). All observations are plotted as red circles.}
    \label{fig:08_robots}
\end{figure*}

The ability for mobile systems to always achieve compatibility can be understood by examining the phase and space dynamics from our theoretical model.
In the fixed base system the only free degree of freedom is $\Delta$ and thus oscillator pairs can only increase or decrease in phase difference (the state evolution in Fig.~\ref{fig:03_model2} is only vertical).
However, when the base is allowed to move in response to collisions the oscillator pairs have a second degree of freedom and the system can evolve through phase change or separation change. 
Because the spatial change between oscillator pairs is only repulsive this emerges as a lateral drift towards larger $\tilde{d}$ in the collision-to-collision state evolution (Fig.~\ref{fig:07_manyoscillators_stats}a).
Anti-phase oscillations are no longer a stable fixed point because the high-impact collisions will drive the oscillators apart until they will settle at the point $(\tilde{d} = 1, \Delta = \pi)$.

To demonstrate that spatial movement inhibits anti-phase oscillation we examined the long time dynamics of an oscillator lattice initialized at a separation distance that leads to anti-phase oscillation in pairs and repetitive collisions in groups ($\tilde{d} = 0.78$).
We varied the the magnitude of collision-induced spatial change, $\beta$, over two orders of magnitude and observed a nearly three order of magnitude increase in the relaxation time  (Fig.~\ref{fig:06_manyoscillators}b).
This power law behavior matches previous simulations and intuition from our model: immobile systems will never relax to collisionless compatibility since $t_{relax} \rightarrow \infty$ as $\beta \rightarrow 0$.
Thus we see a fundamental difference between mobile and fixed-base systems that undulate and interact through collisions, and these results suggest that mobile robots and organisms will always evolve to compatible, collisionless states through contact.

\section{Robots synchronize gaits through collisions}

\label{sec:robot_experiments}

Lastly we examine how groups of mobile undulatory robots synchronize their gaits through contact.
We performed both experiments and simulations with simple three-link robots that have two active servomotors (Dynamixel AX-12, Robotis) controlling joint angles $\alpha_1$, $\alpha_2$, and three rigid links of length 18.65~cm (Fig.~\ref{fig:08_robots}a). 
Such a three-link system is often referred to as ``Purcell's swimmer'' and was originally introduced by E.M. Purcell as a minimal model of low Reynolds swimming \cite{Purcell1977-nk}.  
The three-link robot has been studied extensively in the context of locomotion through fluids \cite{Avron2008-qh,Tam2007-et}, on frictional surfaces \cite{Alben2021-dx, Jing2013-zp}, and within granular media \cite{Hatton2013-yj}. 
In addition three-link robots have been recently used to study the collective behavior of robot groups that exhibit time-dependent oscillatory motion and push each other through contact \cite{Savoie2019-gx, Chvykov2021-jx, Ozkan-Aydin2021-br}. 

In experiment the robot actuators are controlled by continuously sending position commands for the joint angle at a rate of 100~Hz. 
To actuate the robot joints according to the phase-oscillator model using position controlled servos we numerically integrated the oscillator equation used in the motor pair experiments (Eqn.~\ref{eqn:vdp}) solving for the next joint angle at each timestep.
Critically this actuation method required measuring the instantaneous joint angle and joint velocity from the servos and thus incorporates proprioceptive feedback to generate autonomous oscillations, consistent with the direct-drive motors of the previous experiment.

\begin{figure*}[t]
    \centering
    \includegraphics[width=1\linewidth]{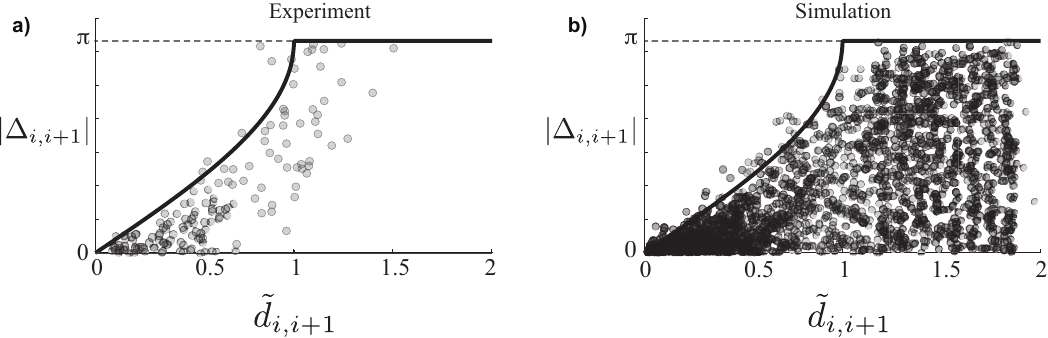}
    \caption{Robots adjust their undulatory phase and lateral distance according to the theoretical model (black curve; Eqn.~\ref{eqn:compatibility}). Results from simulation (n = 210) and experiments (n = 50) at random initial conditions.}
    \label{fig:09_robot_results}
\end{figure*}
 
We simulated the three-link robots using the Project Chrono multibody physics simulation engine \cite{Tasora2016-sq}.
In simulation we directly controlled the torque of the rotational joints consistent with the previous two motor experiment (Eqn.~\ref{eqn:vdp}). 
Contact interactions in the simulation were modeled through short-range repulsive viscoelastic interactions, and we added stokes-drag fluid forces to the robot links according to the method in \cite{Hatton2013-xo} to mimic the damping from friction in experiment. 
In both the experiment and simulation we incorporated methods to enforce a constant phase difference between joints ($\alpha_1 - \alpha_2 \approx \frac{2}{3}\pi$) to produce traveling wave body undulations. 
We modified slightly the actuation equation by adding a coupling term ($\lambda$) between joints $\alpha_1$, $\alpha_2$
\begin{align}
    \tau_{i,j} = -k\left(x_{i,j} + \lambda_j x_{i,\Bar{j}}\right) + \left( c -\mu x_{i,j}^2 \right)\dot{x}_{i,j}
     \label{eqn:limit_cycle_joint_control}
\end{align}
in which subscripts $i = 1, 2, 3, ...$ represents the number of robots and $j = 1, 2$ represents the two joints of robot $i$.
The coupling constants for the two joints were $\lambda_1 = 1.5$ and $\lambda_2 = -0.5$, and the position $x_{i,\Bar{j}}$ refers to the opposite joint of the robot.
The position and torque control methods of the experiment and simulation produced body undulations of the robot with a constant frequency and phase difference (Fig.~\ref{fig:08_robots}b). 
In experiment the frictional interactions between the robot links and ground caused perturbations to the robot joint motion, however this did not affect the synchronization behavior of the robots. 

To observe whether multiple undulatory robots will synchronize their gaits through contact we put groups of four robots within a confined rectangular channel (Fig.~\ref{fig:08_robots}c; supplementary video 2). 
In experiments we only tested configurations where the robots were aligned longitudinally but we tested the effect of longitudinal misalignment in simulation. 
The rectangular channel was 55~cm long and we tested five different widths, $w \in [16, 18, 20, 22, 24]$~cm with 10 trials at each width. 
The experiment began with the robots evenly spaced in the lateral direction and at random initial phases. 
After 30~s (approximately 15 periods of oscillation) we stopped the experiment and measured the final phase difference, $\Delta_{i,i+1}$, and spatial distance, $\tilde{d}_{i, i+1}$, between neighboring robot pairs (Fig.~\ref{fig:08_robots}d).
Increasing the wall width caused both $\Delta_{i,i+1}$ and $\tilde{d}_{i, i+1}$ to increase (Fig.~\ref{fig:08_robots}e).

We performed similar three-link robot synchronization experiments in simulation. 
In addition to simulating the experiments performed with physical robots, we also increased the number of robots and increased the confinement arena size to represent two-dimensional simulations in which robots occupied a rectangular region. 
Qualitatively the one-dimensional and two-dimensional arenas exhibited similar spatial and phase effects, with nearby robots influencing each other in undulatory phase and reaching compatibility.

In all experiments (n = 50) and simulations (n = 210) the three-link robots adjusted their undulatory phase through collisions and the final states were well characterized by the theoretical model of Section~\ref{sec:theory}.
When we examine the nearest neighbor phase difference versus lateral separation we see that all robot-robot interactions lead to phase and distance states that are near, or below the compatibility condition (Eqn.~\ref{eqn:compatibility}; black line in Fig.~\ref{fig:09_robot_results}).
Critically we never observed anti-phase synchronization as we did in the earlier two-joint experiments from section~\ref{sec:experiment1} or the immobile lattice simulations from Section~\ref{sec:1Dsims}. 
The lack of anti-phase behavior is understandable from the mobile simulations in Section~\ref{sec:1Dsims}, when robots collide they push each other away and this spatial repulsion drives them out of contact before they will synchronize to anti-phase.

\begin{figure*}[t]
    \centering
    \includegraphics[width=.95\linewidth]{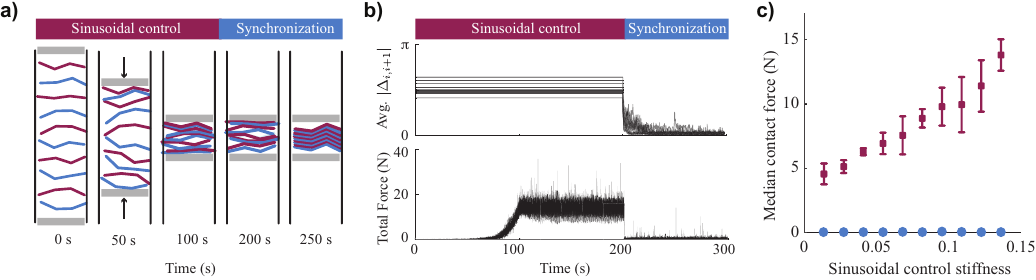}
    \caption{Interaction forces in undulatory groups decrease when robots synchronize gait. a) Ten robots in a rectangular arena oscillate through time-dependent sinusoidal control. The top and bottom walls are slowly brought together resulting in persistent collisions. After 200~s the joint control is switched from time-dependent to phase-oscillator actuation and the robots synchronize. b) The phase difference and contact forces are plotted versus time. During time-dependent sinusoidal control the robot phase differences are maintained and contact forces are large. When phase-oscillator control is switched on the robots synchronize causing the phase difference and contact forces to decrease. Ten replicate experiments are overlaid and force is filtered with a running average of 0.1~s.  c) The median contact forces during steady-state for time-dependent sinusoidal control (purple squares, calculated from 100-200~s) and limit-cycle control (blue circles, calculated from 200-300~s). The forces during time-dependent sinusoidal control linearly increased with the proportional control constant (the effective ``stiffness'').}
    \label{fig:10_sim_forces}
\end{figure*}

The extremely good agreement we observe from both the simulation and experiment with the compatibility model indicates that contact interactions have an important role in collective phase dynamics. 
Initial states outside of compatibility evolved to synchronized movement when spacing was small, and compatible phases at larger spacing.
It is important to note that the mobile robots in simulation and experiment can displace and rotate with respect to each other thus indicating that the phase dynamics model of Section~\ref{sec:theory} is robust to misalignment and natural variation. 
However, it remains to be demonstrated what benefits gait synchronization would have for undulatory collectives. 
In the last section we compare time-dependent actuation in asynchronous group versus undulatory generation through autonomous oscillators that enables synchronization.

\section{Synchronization minimizes contact forces in undulatory groups}

In this last section we seek to determine what is the potential benefit of gait synchronization for collectives. 
There are likely many metrics that could be influenced by synchronization: locomotion energetics and collective sensing for example. 
Here we focus on the interaction forces between robots that occur when in high-density spatial arrangements.

We conducted simulations with groups of ten three-link robots in a confined rectangular volume (Fig.~\ref{fig:10_sim_forces}a). 
Initially the robot joints were actuated through a time-dependent position control signal with fixed frequency and amplitude 
\begin{align}
    \bar{\alpha}_{i,j} = A \sin{\left( \phi_i + \frac{2\pi}{3}(j-1) - 2\pi f t \right)}
     \label{eqn:phase_fixed_joint_control}
\end{align}
in which subscripts $i = 1, 2, 3, ...$ represents the number of robots and $j = 1, 2$ represents the two joints of robot $i$.
The position command was converted to a control torque through a proportional control law $\tau_{i,j} = k~(\alpha_{i,j} - \bar{\alpha}_{i,j})$ where $\alpha_{i,j}$ is the actual joint angle.
The proportionality constant $k$ determines how much torque the actuators exert when the position deviates from the time-dependent sinusoidal commands and can be considered as a controller  ``stiffness''.
We performed simulations across $k \in [0.015, 0.135]~\frac{\mathrm{N m}}{\mathrm{rad.}}$. 
For each control stiffness we performed ten simulations at random initial phases ($\phi_i \in \left[-\pi, ~\pi\right]$ in Eqn.~\ref{eqn:phase_fixed_joint_control}). 
The frequency ($0.8$~Hz.) and amplitude ($0.8$~rad.) were chosen to match the oscillation kinematics when the robots are under limit-cycle control (Eqn.~\ref{eqn:limit_cycle_joint_control}).

To enforce contact and collisions we slowly moved the top and bottom walls inwards towards the arena center.
the rectangular region had a constant width of 0.6~m and at the beginning of the simulation the lateral walls were a distance of 2.0~m apart. 
The width of the lateral walls was decreased at constant velocity from 10~s to 100~s while the robots oscillated, stopping at a lateral width of 0.3~m for the rest of the simulation (Fig.~\ref{fig:10_sim_forces}a).
The robots were controlled through time-dependent sinusoidal actuation during the first 200~s and switched to phase-oscillator control from 200-300~s.

We recorded the oscillatory phase, and the contact forces acting on all robot links during each time step. 
The mean force between robots under the phase-fixed control was significantly larger than that under the phase-oscillator control mode in which synchronization occurred (Fig.~\ref{fig:10_sim_forces}b). 
The interaction forces between the robots in the high-density environment were large because the oscillation phases were incompatible resulting in collisions.
These collisions persisted and repeated because the phase differences were fixed, leading to a fluctuating mean force with constant time-averaged behavior (Fig.~\ref{fig:10_sim_forces}b)
However, when robots were switched to the phase-oscillator control mode, the collisions between robots quickly drove the robot group to synchronization. 
The median contact force was less than 0.1~N during the phase-oscillator control mode indicating a large reduction in contact forces. 
 
This section demonstrates that robots with undulatory phase differences can experience large contact forces as they push against each other.
However when synchronized to the same undulatory phase the collisions reduce to a small and negligible magnitude.
Contact forces between robots can be a significant problem and lead to rapid wear and failure. 
Similar negative consequences are likely to occur in biological collectives where repeated high-force contact can lead to higher energy expenditure and potential injury.

\section{Discussion}

Our results have demonstrated that inelastic collisions between undulatory robots can produce novel phase dynamics such as in-phase and anti-phase synchronization, and compatible oscillations that persist without contact. 
The behavior of larger robot groups tends towards phase compatibility and once achieved the group is effectively decoupled because collisions will no longer occur unless perturbed. 
The compatible state is similar to the ``cohesive'' state originally introduced for the Kuramoto system \cite{Dorfler2011-xy} in which cohesive oscillators remain within a bounded phase difference for all time. 
Compatibility is a beneficial property for undulatory groups because it minimizes the contact forces between individuals and thus likely reduces energetics, fatigue, and damage. 
Critically this beneficial collective behavior emerges naturally from the physics of inelastic contact and simply requires that undulatory motion be generated through an autonomous oscillator so that phases between robots can ``slip'' through interactions. 
In additional simulations and experiments we have demonstrated that this behavior is insensitive to the particular control law that generates undulation.

The coefficient of restitution from inelastic collisions between robots is the lone governing parameter for phase dynamics among these contact-coupled groups. 
Inelastic contact interactions generate a wide array of collective behaviors in driven or active nonlinear systems, such as pattern formation \cite{Aranson2005-ga, Shinbrot1997-ic}, particle aggregation \cite{McNamara1994-rv, Goldman1998-xf, McNamara1994-rv}, and swarming \cite{Grossman2008-dd, Kudrolli2008-ru}. 
However, the ability of repulsive contact interactions to drive attractive phase dynamics in oscillators has not been observed. 
Our phase model is able to explain how phase attraction and repulsion emerges from inelastic collisions. When undulatory systems are in close proximity the collisional interactions between their limit-cycles drive their phase difference to be smaller. However, when the separation distance is large collisions drive the phase difference to grow and generates a stable anti-phase mode. Extending these interactions to an oscillator lattice we have shown that phase repulsion can destroy long range order when the oscillators base is immobilized, while mobile undulatory systems always reach compatibility.

Our inspiration for this study comes from collective movement in worm groups in which body and appendage oscillations may occur in close proximity. 
Recent work has demonstrated that collisional interactions in arrays of cilia can generate synchronization, metachronal wave propagation, and jammed states, dependent on separation \cite{Chelakkot2020-fq}. 
Similarly, recent observations of small worms that swim by laterally oscillating their bodies have illustrated that groups of worms tend to synchronize their oscillatory phase when in close proximity \cite{Yuan2014-xs, Peshkov2021-ji, Quillen2021-zp}. 
Genetic manipulations of these worms illustrated that external sensory responses (exteroception) were not necessary for synchronization, and instead the authors argued that collisional (``steric'') interactions could produce synchronization \cite{Yuan2014-xs}. 
Our results provide a potential explanation for the observed gait synchronization: body oscillations that are governed by internal proprioceptive neural feedback can exhibit emergent synchronization through collisional body interactions alone. 

The system explored in this experiment had appreciable inertial dynamics and momentum transfer through collision. 
However, in the systems we take inspiration from such as small oscillatory organisms in fluids, inertial dynamics are likely not relevant. 
Thus, it is important to consider how these results may apply across inertial and non-inertial active matter systems. We propose that contact-coupled oscillators in both the inertial and non-inertial regimes are captured by the coefficient of restitution in our phase model. 
When $r = 0$, the oscillators do not rebound but instead ``stick'' together which models the non-inertial behavior of oscillators such as cilia and worms in overdamped viscous environments. 
However, for $r > 0$ systems exhibit significant rebounding as they collide which captures the behavior of inertial oscillatory systems and can lead to anti-phase synchronization (Fig.~\ref{fig:03_model2}b). 
The reduction of contact-coupled oscillators to a simple model in which $r$ is the only governing parameter allows us to explore these systems across inertial to non-inertial regimes. 
This will be of interest in future studies and comparisons between model predictions and observations from active matter and swarm robot systems in experiment.

This work has relevance to the field of swarm, and collective robotics where a critical goal is to design distributed control laws that lead to desired, beneficial, emergent behaviors of the group \cite{Yang2018-nt}. 
Recent work in swarm robotics have embraced contact and collisional interactions as a means of coordinating robot group behaviors \cite{Mondada2005-jt, Nemitz2018-fo, Mayya2019-dh, Mayya2017-nj, Mayya2019-hz, Mayya2019-wg, Schmickl2009-qg, Scholz2018-ls, Karimi2020-ix} and other recent work has leveraged collisions \cite{Mulgaonkar2018-nw, Lu2020-yh, Mote2020-vc, Zha2020-cl} for maneuvering individual robots. 
Our work demonstrates that designing appropriate limit-cycles to actuate the rhythmic motion of robots can lead to emergent synchronization and drastically reduce the contact forces. 
Similar principles may be able to be engineered into other robot morphologies and tasks, for example walking robots that collectively push objects through synchronization. 

The coupling of oscillatory dynamics with mobility is an exciting future direction for active matter systems such as biological or robotics swarms. 
Previous work on mobile phase oscillators in which the phase differences can influence motion of the mobile systems have demonstrated novel collective flocking and pattern formation behaviors \cite{OKeeffe2017-fm, Uriu2013-up, Frasca2008-su}. However, there has been little work to consider how the mechanical collisions between oscillating moving individuals drives collective synchrony or motion patterns. 
In recent experiments, three-link ``smarticle'' robots have demonstrated how stochastic interactions among neighboring oscillating robots can drive emergent and controlled collective behavior \cite{Savoie2019-gx}. 
However, currently smarticle systems do not have oscillatory phase dynamics and thus synchronization has not been explored. In our work the oscillator phase is intrinsically tied to the undulatory motion of the robotic joint. 
Thus, phase and motion are explicitly coupled. Future swarm systems that take advantage of the phase dynamics from inelastic collisions may enable emergent synchronization of mobile undulatory robots purely through contact, thus simplifying swarm robot motion control.

\subsection{Acknowledgement}
We acknowledge helpful discussion from Dan Goldman and Paul Umbanhowar. We thank the UCSD department of Mechanical \& Aerospace Engineering for funding support.

\appendix

\section{Derivation of contact map}

\label{sec:phaseoscillator}

In this section we derive the collision-to-collision phase map presented in section~\ref{sec:theory}. To derive this map we have to first consider how to represent the pre-collision phases, $\phi_i^-$, in terms of only the phase difference, $\Delta = \phi_2 - \phi_1$. We seek to solve for the collision phases using only the phase difference between oscillators, $\Delta$. We begin by introducing an intermediate variable $\kappa$ such that
\begin{align}
    \label{eqn:phi1}
    \phi_1 = \kappa - \frac{\Delta}{2} \\
    \label{eqn:phi2}
    \phi_2 = \kappa + \frac{\Delta}{2} 
\end{align}
The collision condition (Eqn.~\ref{eqn:collision_condition} in main text) is 
\begin{align}
    2 \tilde{d} = \cos(\phi_2) - \cos(\phi_1) 
\end{align}
and we expand this into the form
\begin{align}
    2 \tilde{d} &= \cos(\phi_1) - \cos(\phi_2) \\
    &= \cos{(\kappa - \frac{\Delta}{2})} - \cos{(\kappa + \frac{\Delta}{2})}  \\
    &= -2\sin{(\kappa)}\sin{(-\frac{\Delta}{2})}  \\ 
\end{align}
which yields the relationship
\begin{align}
    \label{eqn:kappa}
  \kappa &= \arcsin\left(\tilde{d}\csc\left(\frac{\Delta}{2}\right)\right)
\end{align}
This equation allows us to determine for a given initial $\Delta$ what the individual phases of the oscillators are at collision by substituting $\kappa$ into Equations~\ref{eqn:phi1} \& \ref{eqn:phi2}.

Our goal here is to solve for the return map between collisions as a function of $\Delta$. To do this, we take the following steps: 1) solve for $\phi_1$ and $\phi_2$ at collision from Equations~\ref{eqn:kappa}, \ref{eqn:phi1}, \& \ref{eqn:phi2} and, 2) apply the velocity update rule for inelastic collisions from Equations~\ref{eqn:coeffrest1} \& \ref{eqn:coeffrest2}, 3) determine the post-collision phases for the oscillators from Equation~\ref{eqn:post_collide_phase}. Since $\omega$ is the same between each oscillator, and they evolve independently until colliding, the post-collision phase difference $\Delta^{(n,+)}$, is exactly the same phase difference of the next collision $\Delta^{(n+1,-)}$. We have introduced the superscript notation where the first value indexes the collision, and the $\pm$ denotes whether the value is before ($-$) or after ($+$) the indexed collision. 

\begin{widetext}
\begin{align}
    \Delta^{(n,+)} &= \phi_2^{(n,+)} - \phi_1^{(n,+)} \\
     &= \mathrm{atan} \left[-\frac{\dot{y}_2^{(n,+)}}{y_2^{(n,+)}}\right] - \mathrm{atan} \left[-\frac{\dot{y}_1^{(n,+)}}{y_1^{(n,+)}}\right]\\
    &= \mathrm{atan} \left[\frac{\left(1 - r \right) \dot{y}_1^{(n,-)} +
    \left(1 + r \right) \dot{y}_2^{(n,-)} }{2 y_1^{(n,-)}}\right] -  \mathrm{atan} \left[\frac{\left(1 - r \right) \dot{y}_2^{(n,-)} + \left(1 + r \right) \dot{y}_1^{(n,-)}}{2 y_2^{(n,-)}}\right]  \\
    &= \mathrm{atan} \left[\frac{\dot{y}_1^{(n,-)} + \dot{y}_2^{(n,-)} - r\left(\dot{y}_1^{(n,-)} - \dot{y}_2^{(n,-)}\right)}{2 y_1^{(n,-)}}\right] - \mathrm{atan} \left[\frac{\dot{y}_1^{(n,-)} + \dot{y}_2^{(n,-)} + r\left(\dot{y}_1^{(n,-)} - \dot{y}_2^{(n,-)} \right) }{2 y_2^{(n,-)}}\right]  \\
    &=  \mathrm{atan} \left[\frac{(1 - r)\sin(\kappa + \frac{\Delta^{(n,-)}}{2}) + (1 + r)\sin(\kappa - \frac{\Delta^{(n,-)}}{2})}{2 \cos(\kappa + \frac{\Delta^{(n,-)}}{2})}\right]  \\ \nonumber  & \quad - \mathrm{atan} \left[\frac{(1 - r)\sin(\kappa - \frac{\Delta^{(n,-)}}{2}) + (1 + r)\sin(\kappa + \frac{\Delta^{(n,-)}}{2})}{2 \cos(\kappa - \frac{\Delta^{(n,-)}}{2})}\right] \\
    &= \mathrm{atan} \left[\frac{\sin(\kappa)\cos(\frac{\Delta^{(n,-)}}{2}) - r\cos(\kappa) \sin(\frac{\Delta^{(n,-)}}{2})}{ \cos(\kappa + \frac{\Delta^{(n,-)}}{2})}\right] - \mathrm{atan} \left[\frac{\sin(\kappa)\cos(\frac{\Delta^{(n,-)}}{2}) + r\cos(\kappa) \sin(\frac{\Delta^{(n,-)}}{2})}{ \cos(\kappa - \frac{\Delta^{(n,-)}}{2})}\right]
\end{align}
\end{widetext}
Since $\Delta^{(n,+)} = \Delta^{(n+1,-)}$ we have derived the mapping from the collision phase immediately after the $n$ collision to the phase immediately after the $n+1$ collision. Thus we can drop the $\pm$ superscripts and we arrive at the final collision-to-collision return map
\begin{widetext}
\begin{align}
    \Delta^{(n+1)} = \mathrm{atan} \left[\frac{\sin (\kappa) \cos (\frac{\Delta^{(n)}}{2})- r \cos (\kappa) \sin (\frac{\Delta^{(n)}}{2})}{\cos (\kappa) \cos (\frac{\Delta^{(n)}}{2})-\sin (\kappa) \sin (\frac{\Delta^{(n)}}{2})}\right] - \mathrm{atan} \left[\frac{\sin (\kappa) \cos (\frac{\Delta^{(n)}}{2}) + r \cos (\kappa) \sin (\frac{\Delta^{(n)}}{2})}{\cos (\kappa) \cos (\frac{\Delta^{(n)}}{2}) +\sin (\kappa) \sin (\frac{\Delta^{(n)}}{2})}\right]
\end{align}
\end{widetext}

\section{Experiment details}
\subsection{Motor control and limit-cycle generation}
\label{sec:motor}
Each motor was controlled by an ODrive brushless DC motor controller (ODrive robotics). The ODrive provides closed-loop current control for each motor and we set the maximum current limit to 30~A. The motor current control was performed on a computer in Python. At every update loop the motor current was computed using the following equation:
\begin{align}
i = -k\theta + c\dot{\theta} - \mu\theta^2\dot{\theta} + \beta \text{sgn}(\dot{\theta)}
\end{align}

with the following parameters:
\begin{table}[h]
\centering
\begin{tabular}{lll}
Variable & Motor 0 & Motor 1 \\ \hline
$k$    & 3.9~A/rad  & 3.3~A/rad   \\
$\mu$   & 0.24~A~s/rad$^3$ & 0.24~A~s/rad$^3$  \\
$c$    & 0.009~A~s/rad &  0.009~A~s/rad       \\
$\beta$ & 0.25~A        &  0.25~A
\end{tabular}
\end{table}

The constants were selected so that each motor exhibited limit-cycle oscillations of approximately sinusoidal motion with equal amplitude ($A_1 = 44.4 \pm 0.9$~degrees, and $A_2 = 44.3 \pm 1.6$~degrees) and equal frequency ($\omega_1 = 2.61 \pm 0.04$~Hz, and $\omega_2 = 2.63 \pm 0.03$~Hz). The $\beta$ term in the motor control equation helped overcome the frictional resistance of the motor bearings. Without this term, the motor dynamics exhibited a stable fixed point at $(\theta, \dot{\theta}) = (0, 0)$ with a small region of attraction around this point. 


\subsection{Collision dynamics}
\begin{figure}[t]
  \centering
  \includegraphics[width=1\linewidth]{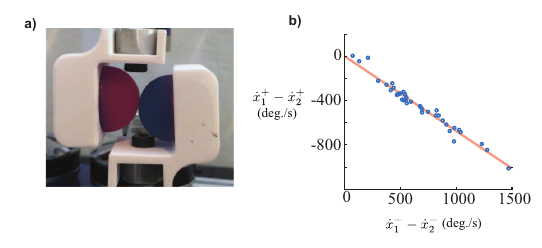}
  \caption{Measurement of coefficient of restitution for experiment. a) Impacting surfaces. b) Coefficient of restitution measurement. Equation is given in text. }
  \label{fig:SI_restitution}
\end{figure}

A rigid robot link of length 9~cm was attached to each motor. The link was waterjet cut from 9.5~mm thick aluminum and rigidly fastened to the motor. A 3D printed adapter was attached to the end of each link which provided an impact surface for the two links to interact with each other (Fig.~\ref{fig:SI_restitution}). The colliding surface was an elastic sphere, a bouncy ball, purchased from a commercial vendor.

To determine the coefficient of restitution of the impacting surfaces we performed a series of experiments. The links were accelerated towards each other at a constant motor current (selected at random between 0 - 2~A) for 200~ms after which the current was set to 0 and the motors and links glided towards each other impacting and rebounding. We measured the motor speed immediately prior to the collision and immediately after the collision and computed the coefficient of restitution using the equation, $\dot{x}_{1}^{+} - \dot{x}_{2}^{+} = -r \left(\dot{x}_{1}^{-} - \dot{x}_{2}^{-}\right)$. We found a coefficient of restitution of $r = 0.67\pm 0.02$.

\subsection{Simulation}
We performed numerical simulations of colliding oscillator pairs and collectives. Simulations were performed in both Matlab and C++ using the library "odeint" and a variable time-step integrator with absolute and relative tolerances of $1\times10^{-6}$. An event detection scheme was used in both simulation environment to detect oscillator collisions. At each collision the numerical integration was halted, the inelastic collision model was implemented, and the integration was re-initialized with the new post-collision state. In the simulations with more than two oscillators simultaneous collisions between more than one oscillator pair were not observed.


 \end{document}